\documentclass[12pt,twoside,letterpaper]{iopart}
\usepackage{times, graphicx, units,mathrsfs,euscript,upgreek,amssymb,pifont}
\usepackage[bookmarks=true,letterpaper=true,colorlinks=true,
linkcolor=blue, citecolor=blue, menucolor=black,pagecolor=blue, filecolor=blue, urlcolor=blue]{hyperref}

\newcommand{\vr}{{\bf r}}
\newcommand{\bk}{{\bf k}}
\newcommand{\bq}{{\bf q}}
\newcommand{\bv}{{\bf v}}
\newcommand{\ebg}{{\epsilon_\mathrm{bg}}}
\newcommand{\ezero}{{\epsilon_0}}
\newcommand{\esub}{{\epsilon_\mathrm{sub}}}
\newcommand{\eeff}{{\epsilon_\mathrm{eff}}}
\newcommand{\Es}{{\epsilon_\mathrm{s}}}

\newcommand{\Phiind}{\Phi_\mathrm{ind}}
\newcommand{\FPhi}{\widetilde{\Phi}}

\newcommand{\FPhiind}{\FPhi_\mathrm{ind}}

\newcommand{\Fs}{F_s}

\begin{document}

\title{Friction force on slow charges moving over supported graphene}

\author{K. F. Allison and Z. L. Mi\v{s}kovi\'{c}}

\address{Dept. of Applied Mathematics, University of Waterloo, Waterloo,
Ontario, Canada N2L 3G1}

\begin{abstract}
We provide a theoretical model that describes the dielectric coupling of a 2D layer of graphene, represented by a
polarization function in the Random Phase Approximation, and a semi-infinite 3D substrate, represented by a surface response
function in a non-local formulation. We concentrate on the role of the dynamic response of the substrate for low-frequency
excitations of the combined graphene-substrate system, which give rise to the stopping force on slowly moving charges above
graphene. A comparison of the dielectric loss function with experimental HREELS data for graphene on a SiC substrate is used
to estimate the damping rate in graphene and to reveal the importance of phonon excitations in an insulating substrate.
A signature of the hybridization between graphene's $\pi$ plasmon and the substrate's phonon is found in the
stopping force. A friction coefficient that is calculated for slow charges moving above graphene on a metallic substrate
shows an interplay between the low-energy single-particle excitations in both systems.
\end{abstract}
 \submitto{\NT}

\section{Introduction}

After the initial wave of intense studies of electronic and transport properties of graphene \cite{Castro_2009,Adam_2009}, the
focus of research seems to be turning to graphene's interactions with a dielectric environment
\cite{Adam_2009,Ponomarenko_2009}. While it has been recognized early on that the presence of charged impurities in the material
surrounding graphene is a likely cause of its peculiar minimum conductivity \cite{Tan_2007}, much attention has also been paid
to the dielectric screening of these impurities by either their host material or various high-kappa dielectrics
\cite{Jang_2008}. Besides an obvious interest in SiO$_2$ as the most common substrate for graphene that is produced by the exfoliating
technique \cite{Romero_2008,Sonde_2009}, interactions of epitaxially-grown graphene on a SiC substrate have also attracted
considerable attention \cite{Zhou_2007,Rotenberg_2008}. Most recently, several reports have appeared studying the properties of
graphene on metallic substrates \cite{Wintterlin_2009,Khomyakov_2009}, as well as graphene's interaction with organic solvents and
ionic solutions \cite{Chen_2009a,Chen_2009b}.

In view of these developments and possible extensions of the recent Electron Energy Loss Spectroscopy (EELS) studies of
free-standing graphene structures \cite{Kramberger_2008,Eberlein_2008} that would include the presence of a substrate,
it is of interest to study the interaction of graphene with externally-moving charged particles in a dielectric environment.
Although it was recently shown that a dielectric environment can exert strong effects on the plasmon excitations of carbon nanotubes due to coupling
of their $\sigma+\pi$ and $\pi$ plasmons with high frequency modes in the nearby material \cite{Mowbray_2006}, in this paper we
limit ourselves to the effects of a substrate on slow-moving external charges such that
only the low-energy excitations of graphene's $\pi$ electron bands occur. From an applied point of view, this
problem is highly relevant to a recent experiment that used a High-Resolution EELS (HREELS) technique to probe the
momentum-resolved low-frequency plasmon excitations of doped graphene grown on a SiC substrate \cite{Liu_2008}. Moreover,
interactions with slow heavy particles are relevant to studies of intercalation of alkali-metal atoms
\cite{Khantha_2004}, friction forces on migrating atoms and molecules moving near graphene \cite{Tomassone_1997,Dedkov_2002},
and monitoring of ion flow in an aqueous solution adjacent to carbon nano-structures \cite{Ghosh_2003}.

From a theoretical point of view, the interaction of graphene with slow external charges can be described by the recently developed dielectric
function in the Random Phase Approximation (RPA), in which graphene's $\pi$ electron bands are treated in the approximation of linearized
electron energy dispersion, giving rise to the picture of massless Dirac fermions (MDF) in two dimensions (2D) \cite{Wunsch_2006,Hwang_2007}.
This theoretical model captures all of the physical processes relevant to such interactions, including graphene's 2D $\pi$-plasmon mode and both
the intra-band and inter-band single particle excitations (SPEs). Furthermore, the MDF-RPA dielectric function for graphene takes into account
the strong dependence of these excitation modes on the level of doping in graphene, which often results from charge transfers between graphene
and the substrate \cite{Romero_2008,Liu_2008,Khomyakov_2009} or may be induced by a gate potential. However, all RPA dielectric functions suffer
two major shortcomings: they neglect the local-field effects (LFE) due to electron correlations, and they have an inherent assumption of an
infinite lifetime of electron excitations. Only one of these shortcomings can be qualitatively corrected in the MDF-RPA dielectric function for
graphene at a time, e.g., by using either the Hubbard approximation (HA) for the LFE \cite{Adam_2009} or by introducing a finite decay rate,
$\gamma$, using Mermin's procedure \cite{Mermin_1970,Qaiumzadeh_2008}.

With respect to graphene's coupling with a substrate, we note that few theoretical studies have gone beyond
treating the substrate in a static mode and most assume a vanishing graphene-substrate gap, with the exception of recent studies
of the remote scattering of charge carriers in graphene on surface phonon modes of a strongly polar substrate
\cite{Fischetti_2001,Fratini_2008,Konar_2009}. Therefore, in this work we first provide a detailed
description of graphene's coupling with a substrate by combining the LFE or Mermin-corrected MDF-RPA dielectric function
for graphene with the surface response function for a substrate, which is described by a bulk dielectric function that is allowed to depend
on both the frequency and the three-dimensional (3D) wavenumber. We note that the dependence of the substrate's dielectric function on the
wavenumber is often neglected, giving rise to the local approximation for its surface response function that is appropriate
for describing cases such as the non-dispersing Fuchs-Kliever surface phonons in strongly polar substrates
\cite{Fratini_2008,Konar_2009}. However, when the substrate's bulk dielectric function depends strongly on the 3D wavenumber,
such as in the case of a metal, one has to resort to the so-called Specular Reflection Model (SRM) \cite{Ritchie_1966} in order to
obtain a surface response function that takes into account the spatial dispersion of the substrate's excitation modes, and hence gives
rise to a non-local description of the substrate.

While such a formulation of the graphene-substrate interaction allows for the possibility of studying the hybridization of
collective modes in the two systems \cite{Mowbray_2006}, we focus on analyzing the effects of the graphene-substrate
dielectric coupling on the energy losses of a slow external charge moving parallel to graphene.
It has been experimentally confirmed that slow ions can lose their kinetic energy during grazing
scattering from an insulating surface due to excitations of optical phonons in the substrate \cite{Borisov_1999}, and it would
therefore be interesting to explore how such a process would change in the presence of graphene and the possibility of
simultaneous excitations of graphene's $\pi$ plasmon and its intra-band SPEs. Moreover, techniques
such as HREELS can be used to probe the hybridization of the substrate's phonon modes with graphene's $\pi$ plasmon under
controlled doping \cite{Fischetti_2001,Konar_2009}, and further explore the role of the nearby SPE continuum as a
region in which such hybridized modes can undergo Landau damping. Metallic substrates, on the other hand, have attracted attention only
recently as a possibly interesting dielectric environment for graphene \cite{Wintterlin_2009,Khomyakov_2009}. It has been
recognized for many years that metal surfaces can exert strong dissipative and conservative (image) forces on moving external
charges, especially in the grazing scattering geometry \cite{Winter_2002}. In particular, low-energy SPEs in a metal have been
shown to give rise to strong friction on particles moving slowly over metal surfaces
\cite{Dedkov_2002,Nunez_1980,Echenique_2001}. It is therefore tempting to explore how this concept of friction would be affected
by the presence of graphene on a metal substrate, resulting in an overlap of the metal's SPEs and graphene's intra-band SPEs.
Similar effects have been explored using a non-local description for a Cu overlayer on a Si substrate \cite{Kwei_2003}.

Consequently, we believe that studying both the loss function of the graphene-substrate system and the stopping force on an
external moving charge can be used to explore the effects of substrate's dynamic response. There are many model
parameters that may be relevant to the description of such processes. In our previous work \cite{Allison_2009}, we have
concentrated on amending the MDF-RPA dielectric function for graphene with a constant damping rate, $\gamma$, as well as on analyzing the effects
of the equilibrium charge carrier density in graphene, $n$, and the graphene-substrate gap, $h$, for a
substrate treated in the static mode. In this work, we consider further refinements to the MDF-RPA dielectric
function for graphene, including a dispersing damping rate and the LFE correction. However, our main focus is on the
inclusion of the dynamic and non-local properties of a substrate's dielectric response, for which we consider two specific examples of
low-frequency excitations: a SiC substrate with a pronounced transverse optical (TO) phonon, and an Al substrate with a
well-defined spectrum of low-energy SPEs. In doing so, we reduce the parameter space of our model by considering only
two equilibrium charge densities: $n=0$, characterizing intrinsic or undoped graphene, and $n=n_0$, where $n_0=10^{13}$
cm$^{-2}$, as a value very close to graphene's charge carrier density for both the SiC \cite{Liu_2008} and Al \cite{Khomyakov_2009}
substrates. We also take $h=1$ \AA\ as a reasonable estimate for the graphene-substrate gap
\cite{Khomyakov_2009,Allison_2009}.

After outlining the basic theory in the following section, we present results comparing our model with the HREELS
experiment of Liu \textit{et al.} \cite{Liu_2008}. We then consider the effects of various model parameters on the stopping force for
free graphene, followed by examples of the stopping force for graphene on a SiC substrate and the friction
coefficient for graphene on an Al substrate. A brief discussion of these results is provided in the concluding section.
We use Gaussian electrostatic units throughout this paper unless otherwise stated.

\section{Theory}

We use a Cartesian coordinate system with coordinates $\{\vr,z\}$, where $\vr=\{x,y\}$, and assume that graphene is placed in
the $z=0$ plane. A semi-infinite substrate with a bulk dielectric function $\esub(\bk,\omega)$ is assumed to occupy the region
$z\le -h$ underneath graphene, while the region $z>-h$ is assumed to be a vacuum or air \cite{Radovic_2008}. The
3D wave vector describing the excitation modes in the substrate can be decomposed as $\bk=\{\bq,k_z\}$,
where $\bq=\{k_x,k_y\}$ are the components parallel to the substrate surface and $k_z$ is the component perpendicular to the substrate surface.
We further assume that an external charge with density $\rho_\mathrm{ext}(\vr,z,t)$ moves along a classical trajectory
localized to the region $z>0$ above graphene. Following Ref.\cite{Radovic_2008}, we can express the induced potential
$\Phiind(\vr,z,t)$ in the region above graphene by using the 2D Fourier transform with respect to the surface components
($\vr\rightarrow\bq$) and with respect to time ($t\rightarrow\omega$) as
\begin{eqnarray}
\FPhiind^>({\bq},z,\omega)= \left[\frac{1}{\epsilon(\bq,\omega)}-1\right]\,\frac{2\pi}{q}\,S(\bq,\omega)\,\mathrm{e}^{-qz} ,
\label{Phiind}
\end{eqnarray}
where $\epsilon(\bq,\omega)$ is the dielectric function of the combined graphene-substrate system and
\begin{eqnarray}
S(\bq,\omega)=\int\limits_{-\infty}^\infty dt\,\mathrm{e}^{i\omega t}\int d^2\vr\,\mathrm{e}^{-i\bq\cdot\vr}\int\limits_0^\infty
dz\,\mathrm{e}^{-qz}\,\rho_\mathrm{ext}(\vr,z,t)
 \label{Structure}
\end{eqnarray}
is the structure factor of the external charge distribution. The dielectric function of the system can be written as
\begin{eqnarray}
\epsilon(\bq,\omega)=\ebg(q,\omega)+V_\mathrm{C}(q)\Pi(q,\omega), \label{eps}
\end{eqnarray}
where $\ebg(q,\omega)$ is the effective background dielectric function of the substrate, $V_\mathrm{C}(q)=2\pi
e^2/q$ is the Fourier transform of the 2D Coulomb interaction, and $\Pi(q,\omega)$ is the polarization
function for free graphene.

Assuming that the substrate is isotropic so that its dielectric function only depends on the magnitude of the wave vector,
$k=\sqrt{q^2+k_z^2}$, one can use the SRM  to handle the electrostatic boundary condition on the
substrate's surface \cite{Ritchie_1966}. Combining the steps of the derivation outlined in Ref.\ \cite{Denton_1998} with those in
\cite{Radovic_2008}, we obtain
\begin{equation}
\label{eq:ebg}
\ebg(q,\omega)=\left[1-\frac{\Es(q,\omega)-1}{\Es(q,\omega)+1}\,\mathrm{e}^{-2qh}\right]^{-1},
\end{equation}
where $\Es(q,\omega)$ is the effective surface dielectric function of the substrate, defined by
\begin{eqnarray}
\Es(q,\omega)=\left[\frac{q}{\pi}\int_{-\infty}^\infty\frac{dk_z}{k^2\esub(k,\omega)}\right]^{-1}. \label{Es}
\end{eqnarray}
The expression $(\Es-1)/(\Es+1)$ in Eq.\ (\ref{eq:ebg}) represents a non-local surface response function for a substrate
that exhibits strong dispersion, but in cases where the dispersion in the substrate can be neglected so that
$\esub(k,\omega)\approx\esub(\omega)$, one recovers the local approximation $\Es(q,\omega)\approx\esub(\omega)$.

As an example of a non-local surface dielectric function, one may consider the low-frequency expansion of Lindhard's
dielectric function, \cite{Dedkov_2002,Nunez_1980,Echenique_2001}
\begin{eqnarray}
\esub(k,\omega)\approx 1+\frac{K^2_\mathrm{TF}}{k^2}+i\frac{\pi\omega}{kV_F}\,H\!\!\left(2K_F-k\right), \label{Lindhard}
\end{eqnarray}
where $K_\mathrm{TF}=\sqrt{3}\Omega_p/V_F$ is the Thomas-Fermi (TF) inverse screening length of a 3D electron gas having
a volume density $N$, $\Omega_p=\sqrt{4\pi e^2 N/m_e}$ is its bulk plasma frequency, $K_F\equiv m_eV_F/\hbar=(3\pi^2
N)^{1/3}$ is its Fermi wavenumber, and $m_e$ is the mass of an electron. This dielectric function describes the low-energy SPEs of a metal
subject to the cut-off $k\le 2K_F$, as implied by the Heaviside function $H\left(2K_F-k\right)$ in Eq.\
(\ref{Lindhard}). We note that Eq.\ (\ref{Lindhard}) has often been used in Eq.\ (\ref{Es}) to model various
surface precesses on metals \cite{Dedkov_2002,Nunez_1980,Echenique_2001}.

On the other hand, the local approximation may be suitable for cases such as a strongly polar insulating substrate characterized by
non-dispersing TO phonon modes with frequencies $\omega_\mathrm{TOi}$ and damping rates
$\gamma_\mathrm{TOi}$, in which case one may use a dielectric function of the form \cite{Fischetti_2001,Fratini_2008,Konar_2009}
\begin{equation}
\label{eq:esub} \esub(\omega)=\epsilon_\infty+\sum_i
\frac{f_i\,\omega_\mathrm{TOi}^2}{\omega_\mathrm{TOi}^2-\omega\!\left(\omega+i\gamma_\mathrm{TOi}\right)},
\end{equation}
where $\epsilon_\infty=\lim_{\omega\rightarrow\infty}\esub(\omega)$ is the high-frequency dielectric constant of the substrate
and the oscillator strengths, $f_i$, satisfy the relation $\sum_if_i=\ezero-\epsilon_\infty$, where $\ezero=\esub(0)$ is the
static dielectric constant of the substrate \cite{Fischetti_2001,Fratini_2008}.

We note that in the literature on graphene it is common to assume a zero gap ($h=0$) and to describe the substrate by only the static
dielectric constant \cite{Adam_2009,Wunsch_2006,Hwang_2007}, for which Eq.\ (\ref{eq:ebg}) gives an effective background
dielectric constant $\ebg(q,\omega)\rightarrow\eeff=(\ezero+1)/2$. In this case, a simple description of the screening of
electron-electron interactions in graphene can be quantified by the Wigner-Seitz radius, $r_s=e^2/(\eeff\hbar v_F)$
\cite{Hwang_2007}, where $v_F\approx c/300$ is the Fermi speed of graphene and $c$ is the speed of light in vacuum. We note that the
limiting case of free graphene ($\ebg(q,\omega)=1$) can be obtained by letting $h\rightarrow\infty$ in Eq.\
(\ref{eq:ebg}) or by setting $\ezero=1$ in the zero gap case with a static substrate, but the
focus of this work is on the effects of substrate including a finite gap, non-local effects, and its dynamic
response.

For the dielectric response of graphene in Eq.\ (\ref{eps}), we use the polarization function for graphene's
$\pi$ electron excitations in the MDF-RPA \cite{Wunsch_2006,Hwang_2007}, $\Pi(q,\omega)$, as well as the corrections to
the polarization function for the LFE and the finite lifetime of the excitation modes of charge carriers in graphene.
The LFE are described in the Hubbard approximation (HA) by replacing the MDF-RPA polarization function $\Pi(q,\omega)$ with
$\Pi_\mathrm{LFE}(q,\omega)=\left[1-G(q)\right]\Pi(q,\omega)$, where $G(q)=(q/g_\mathrm{d})/\sqrt{q^2+k_F^2}$ and
$g_\mathrm{d}=4$ is the degeneracy factor of graphene's $\pi$ electrons \cite{Adam_2009}. The finite lifetime of the excitation modes
of charge carriers in graphene is treated by
introducing a finite damping rate, $\gamma$, in the MDF-RPA polarization function through Mermin's procedure
\cite{Mermin_1970,Qaiumzadeh_2008}, whereby one replaces $\Pi(q,\omega)$ with
\begin{equation}
\label{Pi_M} \Pi_M(q,\omega,\gamma)=\frac{\Pi(q,\omega+i\gamma)}{1-\displaystyle{\frac{i\gamma}{\omega+i\gamma}\left[1-
\frac{\Pi(q,\omega+i\gamma)}{\Pi_s(q)}\right]}}, \label{Mermin}
\end{equation}
where the static limit of the RPA polarization, $\Pi_s(q)=\lim_{\omega\rightarrow0}\Pi(q,\omega)$, is given elsewhere
\cite{Wunsch_2006,Hwang_2007}.

In the case of a point charge $Ze$ moving parallel to graphene with velocity $\bv$ and at a fixed distance $z_0>0$, we obtain
$S(\bq,\omega)=2\pi Ze\,\delta(\omega-\bq\cdot\bv)\,\mathrm{e}^{-qz_0}$ from Eq.\ (\ref{Structure}). Using this expression in the inverse
Fourier transform of Eq.\ (\ref{Phiind}), we can evaluate the stopping force on the point charge $Ze$ from the definition
\begin{eqnarray}
F_s=-Ze\,{\bf \hat{v}}\cdot\nabla \left.\Phiind^>(\vr,z,t)\right\vert_{\vr\!=\!\bv t,z\!=\!z_0},
 \label{Fs}
\end{eqnarray}
where ${\bf \hat{v}}=\bv/v$, giving \cite{Radovic_2008}
\begin{eqnarray}
\Fs=\frac{2}{\pi}\frac{Z^2e^2}{v}\int_0^\infty
dq\,\mathrm{e}^{-2qz_0}\int_0^{qv}d\omega\,\frac{\omega}{\sqrt{q^2v^2-\omega^2}}\,\,
\Im\!\left[\frac{1}{\epsilon(q,\omega)}\right]. \label{stopping2}
\end{eqnarray}
We note that $\Im\!\left[-1/\epsilon(q,\omega)\right]$ represents the loss function of the system that,
apart from a kinematic prefactor describing the scattering event, is directly accessible
in the HREELS experiment \cite{Liu_2008,Ibach_1982}.

Finally, for cases in which the loss function in Eq.\ (\ref{stopping2}) can be expanded to the first order in frequency as
$\Im\!\left[-1/\epsilon(q,\omega)\right]\approx\omega\mathcal{F}(q)$,
one can define a friction coefficient $\eta$ for a sufficiently slow particle moving parallel
to graphene through the equation $\Fs\approx-\eta v$. In such cases, the friction coefficient is given by
\begin{eqnarray}
\eta=\frac{1}{2}Z^2e^2\int_0^\infty dq\,q^2\,\mathrm{e}^{-2qz_0}\mathcal{F}(q). \label{eta}
\end{eqnarray}
For example, this procedure has been performed using Eqs.\ (\ref{Es}) and (\ref{Lindhard}) to obtain the friction coefficient for
slow particles on metal surfaces \cite{Dedkov_2002,Nunez_1980}. A similar approach can also be used for graphene in the limit
of vanishing damping rate, giving
\begin{eqnarray}
\Pi(q,\omega)\approx \Pi_s(q)+\frac{i\omega}{\pi\hbar v_F^2}\,H\!\!\left(2k_F-q\right)\sqrt{\left(\frac{2k_F}{q}\right)^2-1},
\label{loss}
\end{eqnarray}
where $k_F=\sqrt{\pi n}$ is graphene's Fermi wavenumber. A different but analogous expression can be obtained from Eq.\
(\ref{Mermin}) in the case of a finite damping rate \cite{Allison_2009}.

\section{Results}

\subsection{Comparison with HREELS experiment}

It has been known for some time that the damping rate $\gamma$ of elementary excitations in a many-electron system can
strongly influence the friction forces on slow particles \cite{Dedkov_2002,Nunez_1980}. However, the exact value of the damping rate is
generally difficult to measure, and in the case of graphene is presently unknown. Nevertheless, $\gamma$ is often used as an empirical
parameter in the RPA response function modified by Mermin's procedure \cite{Mermin_1970,Qaiumzadeh_2008}. Therefore, to obtain a
reasonable estimate for the value of $\gamma$ for graphene, we compare the experimental data of Liu \emph{et al}.\
\cite{Liu_2008} for the HREELS spectra of doped graphene on an SiC substrate with the loss function
$\Im\!\left[-1/\epsilon(q,\omega)\right]$ obtained from Eq.\ (\ref{eps}) by using the Mermin polarization function for graphene,
which is given in Eq.\ (\ref{Mermin}). In addition to constant values of $\gamma$, we also explore a damping rate with a linear
dispersion of the form $\gamma=v_cq$, where $v_c$ is a constant speed. The gap height $h$ is also
treated as an empirical parameter with a reasonable value of 1 \AA. We note that our comparison with the HREELS experiment
is only tentative because we do not include the low-frequency contributions from scattering kinematics or
temperature effects \cite{Ibach_1982}. Nevertheless, we expect that the loss function obtained from the Mermin
polarization function with a suitable choice of $\gamma$ will give the correct order of magnitude for spectral widths at frequencies
that are not too low.

The effects of a SiC substrate on graphene are not as well understood as those of a SiO$_2$ substrate, but without entering the
current debate \cite{Zhou_2007,Rotenberg_2008} we simply neglect any changes in graphene's $\pi$-band structure that may result
from a hybridization of its $\pi$ orbitals with the substrate. In addition to treating the SiC substrate in the static mode with dielectric
constant $\epsilon_0$ = 9.7, we reproduce some of the low-frequency features from the HREELS spectra
\cite{Liu_2008} in a qualitative manner by including the dominant TO phonon mode of SiC in Eq.\ (\ref{eq:esub}) with
eigen-frequency $\omega_\mathrm{TO}$ = 97 meV \cite{Fratini_2008}, damping rate $\gamma_\mathrm{TO}$ = 10 meV \cite{Nienhaus_1995},
and high-frequency dielectric constant $\epsilon_\infty$ = 6.5 \cite{Fratini_2008}. Also, by including the TO phonon in our
comparison with the HREELS data, we gain some insight into the dynamic response of an insulating substrate prior
to using it in calculations of the stopping force on a moving charge.

In Fig.\ 1, we display a comparison of the HREELS data \cite{Liu_2008} for wavenumbers ranging from 0.008 to 0.126
\AA$^{-1}$ with the loss function obtained from the MDF-RPA polarization function with $\gamma=0$ (solid lines) and
the loss functions obtained from the Mermin polarization function with
$\gamma=v_Fk_0$ (dashed lines) and $\gamma=v_Fq$ (dotted lines), where $k_0=\sqrt{\pi n_0}$ and $n_0=10^{13}$ cm$^{-2}$.
We also show cases for which the SiC phonon is included in the model (thick lines) and for which the SiC substrate is treated in the static mode
(thin lines). The equilibrium charge carrier density in graphene is set at $n=1.9\times 10^{13}$ cm$^{-2}$ (hence $k_F\approx$
0.08 \AA$^{-1}$, giving the Fermi energy $\varepsilon_F=\hbar v_Fk_F\approx 570$ meV) to match experimental conditions
\cite{Liu_2008}. Note that since the HREELS data is scaled to arbitrary units of intensity, the loss functions with finite
$\gamma$ are scaled so that their maximum peak heights coincide with those from the experiment.

From Fig.\ 1, it can be seen that the best fit to the data can be obtained with the constant value $\gamma=v_Fk_0$, while the best fit of the
linear damping rate, $\gamma=v_Fq$, tends to underestimate the widths of the experimental spectra at long wavelengths. Moreover, it appears that
the inclusion of the substrate's TO phonon hints at the experimentally observed non-dispersing feature at approximately 100 meV. For the
$\gamma=0$ case, the two sets of narrow peaks indicate the position of graphene's $\pi$ plasmon at long wavelengths before it crosses into the
inter-band SPEs continuum at $q\approx0.07$ \AA$^{-1}$ and becomes broadened from Landau damping. Interestingly, the plasmon peak is broader in
the case with the substrate phonon than in the case of a static substrate. This is a consequence of the hybridization of the substrate phonon
with damping rate $\gamma_\mathrm{TO}$ = 10 meV and graphene's $\pi$ plasmon with a vanishing damping rate. Therefore, using the known value of
$\gamma_\mathrm{TO}$ = 10 meV as a ``marker'', one can surmise that the peaks shown in Fig.\ 1 with thick solid lines at long wavelengths are,
in fact, the substrate phonon that is promoted due to an avoided crossing with the $\pi$ plasmon \cite{Mowbray_2006}. This idea will be
discussed in more detail shortly.

Since the HREELS experiment \cite{Liu_2008} has been designed to determine the plasmon dispersion in graphene on a SiC substrate, in Fig.\ 2
we show the peak positions deduced from the experimental spectra and from
the various theoretical models displayed in Fig.\ 1. In addition, we show
the peak positions resulting from the polarization function for graphene with the LFE correction,
$\Pi_\mathrm{LFE}(q,\omega)$, in the $\gamma=0$ case (dash-dot lines) \cite{Adam_2009}. While it is clear that the LFE correction is unable to reproduce
the widths of the experimental spectra, Fig.\ 2 also demonstrates that its effects on the plasmon dispersion are hardly
discernible from the case of the MDF-RPA polarization with $\gamma=0$. Fig.\ 2 further confirms that the case
$\gamma=v_Fk_0$ provides the best fit to the experimental plasmon dispersion. For the static
substrate case (thin lines), one notices that the peak positions approach the value $\omega=0$ as $q\rightarrow 0$ in a manner that
depends on damping rate, so that a typical 2D plasmon with $\omega\sim\sqrt{q}$ is produced for $\gamma=0$ and $\gamma=v_cq$,
while a constant $\gamma$ appears to give rise to a quasi-acoustic relation $\omega\sim q$ when $q\rightarrow 0$.

Finally, in all models of the plasmon peaks, the inclusion of the substrate phonon (thick lines) appears to capture the tendency of
the experimental data to approach a finite frequency rather than vanish as $q\rightarrow 0$. A closer
inspection shows that the frequency in question is near the value of 116 meV, which corresponds to the Fuchs-Kliever (FK)
surface phonon at the frequency
$\omega_\mathrm{FK}=\omega_\mathrm{TO}\sqrt{\left(\epsilon_0+1\right)/\left(\epsilon_\infty+1\right)}$
\cite{Fratini_2008,Nienhaus_1995}. This behavior is commensurate with the concept of a phonon-plasmon hybridization
\cite{Fischetti_2001}, which is illustrated in Fig.\ 2 by two thick solid lines showing the zeroes of the dielectric function
for the MDF-RPA polarization function with $\gamma=0$. These two lines display an avoided
crossing near the point where the thin solid line (representing graphene's $\pi$ plasmon dispersion for a static substrate)
would cross the FK phonon frequency $\omega_\mathrm{FK}\approx$ 116 meV. One can see from Fig.\ 2 that the substrate phonon
frequency starts at $\omega_\mathrm{FK}$ for $q=0$ and is promoted to higher frequencies as $q$ increases, following a
dispersion curve similar to that of graphene's $\pi$ plasmon in the case of a static substrate. This promoted phonon dispersion
enters the inter-band SPE continuum at $\omega>v_F(2k_F-q)$ with $\omega>v_Fq$. On the other hand, graphene's plasmon starts
from the origin of the $(q,\omega)$ plane and initially follows the standard long wavelength plasmon dispersion
$\omega_p(q)=v_F\sqrt{2r_sk_Fq}$ \cite{Wunsch_2006,Hwang_2007,Radovic_2008}, but quickly levels to the TO phonon
frequency $\omega_\mathrm{TO}$ = 97 meV as $q$ increases. As a consequence of this avoided crossing \cite{Mowbray_2006}, the
plasmon branch is pushed into the intra-band SPEs continuum of graphene at $\omega<v_Fq$, where it broadens
after crossing the point $q_c\approx\omega_\mathrm{TO}/v_F\approx$ 0.015 \AA$^{-1}$ from Landau damping. We note that this complicated
scenario is depicted in Fig.\ 2 for $n=1.9\times 10^{13}$ cm$^{-2}$, which is the spontaneous doping density of graphene on an
SiC substrate \cite{Liu_2008}, but by changing graphene's charge carrier density with a gate potential, for example, one can alter the picture of
the plasmon-phonon hybridization quite significantly.

\subsection{Stopping force with a dynamic substrate}

After the discussion on the low-frequency excitations in graphene on an insulating substrate with a prominent TO
phonon, we now explore how the various model parameters of this system affect the stopping force on a point charge, defined in
Eq.\ (\ref{stopping2}). Given the complexity of the system, we first analyze the case of free graphene by letting
$h\rightarrow\infty$. In Fig.\ 3, we show the velocity dependence of the stopping force on a proton ($Z=1$) moving a
distance $z_0$ = 10 \AA\ away from graphene for the cases of the MDF-RPA polarization function with $\gamma=0$ (solid lines), the Mermin
polarization function with $\gamma=v_Fk_0$ (dashed lines) and $\gamma=v_Fq$ (dotted lines), and
the LFE polarization function with $\gamma=0$ (dash-dot lines), where $k_0=\sqrt{\pi n_0}$ and $n_0=10^{13}$ cm$^{-2}$.
Results are provided for intrinsic graphene
($n=0$) and for graphene with a charge carrier density $n=n_0$. From Fig.\ 3, one can see that all models give qualitatively similar results
at high speeds, but at low speeds there are significant differences in the behaviour of the stopping force due to variations in the
charge carrier density $n$ and the damping rate $\gamma$.
In particular, a threshold velocity at $v=v_F$ exists for the cases of intrinsic graphene with
$\gamma=0$ due to the lack of intra-band SPEs and the presence of inter-band SPEs
in the $\omega>v_Fq$ region of the $(q,\omega)$ plane \cite{Wunsch_2006,Hwang_2007}. Furthermore,
all cases with $\gamma>0$ or $n>0$ result in a linear dependence of the stopping force on the
particle speed, which can be described using the concept of friction. The slopes of these curves do not appear to be
affected significantly by the choice of a constant or linear $\gamma$, but the slopes do
depend significantly on the density, as discussed previously \cite{Allison_2009}. It also appears that the LFE correction to
the MDF-RPA polarization function has a small effect on the stopping force for speeds in the range $v<v_F$. Therefore, we may conclude
that the major influences on the polarization of graphene by an external charge are
determined by the equilibrium charge carrier density $n$ and the damping rate $\gamma$, which can be best described by a constant value.

We now focus on analyzing the effects of the TO phonon in a SiC substrate on the stopping force of a proton moving a distance
$z_0$ = 10 \AA\ away from graphene with a graphene-substrate gap height of $h=1$ \AA.
Given our results for free graphene, in Fig.\ 4 we present the stopping force for the damping rates $\gamma=0$ and
$\gamma=v_Fk_0$, as well as for the two charge carrier densities $n=0$ and $n=n_0=10^{13}$ cm$^{-2}$.
In all cases, the
inclusion of the phonon (thick lines) results in a surprisingly large increase of the stopping force for speeds in the range
$v>2v_F$ when compared with the case of a static substrate (thin lines). This is due to the fact that as the speed
increases, the integration range $0<\omega<vq$ in Eq.\ (\ref{stopping2}) probes a larger portion of the promoted
phonon dispersion curve shown in Fig.\ 2, which approaches a finite frequency rather than vanishing as $q\rightarrow 0$. However,
we are more interested in the stopping force at low speeds ($v<v_F$) for which the phonon-plasmon hybridization produces some peculiar
effects. We first analyze the $\gamma=0$ case in Fig.\ 4(a). For intrinsic graphene, which does not
support a plasmon mode, one can see from the inset that the inclusion of the substrate phonon gives rise to
a small but non-vanishing stopping force in the subthreshold region due entirely to the non-dispersing FK phonon at
frequency $\omega_\mathrm{FK}\approx$ 116 meV. For the finite density $n=n_0$,
the inset in Fig.\ 4(a) shows that the linearly increasing stopping force in the static substrate case is augmented by a contribution from the
original $\pi$ plasmon, which is pushed into the intra-band SPE region due to an avoided crossing with the substrate phonon and traverses this
region at the constant frequency $\omega_\mathrm{TO}\approx$ 97 meV without dispersion but with broadening from Landau damping. It is important
to note that the inclusion of the TO phonon destroys the linear velocity dependence of the stopping force at low speeds, making the concept of
friction inappropriate for the $\gamma=0$ case. Moving to the case of $\gamma = v_F k_0$, which is shown in Fig.\ 4(b), one notices that the
large peak in the magnitude of the stopping force has been broadened and suppressed. Furthermore, although the effects of the substrate phonon
are somewhat softened at low speeds, they still result in a non-linear dependence of the stopping force on the velocity for both $n=0$ and
$n=n_0$.

Finally, we analyze the stopping force of a slow charge moving parallel to a layer of graphene on an Al surface. From a first-principles
calculation, it has been shown that graphene's $\pi$ electron band is not destroyed by a hybridization with aluminum, leaving its
elementary excitation modes intact \cite{Khomyakov_2009}. However, aluminum is close to an ideal metal with a volume
density of quasi-free electrons given by $N=3/(4\pi R_s^3)$, where $R_s\approx 2.07\,a_B$ and $a_B$ is the Bohr radius,
and supports a high-frequency bulk plasmon as well as a well-defined continuum of low-energy SPEs. Without addressing the problem of a
hybridization of graphene's $\pi$ plasmon and aluminum's surface plasmon, we explore the interplay of the intra-band
SPEs in graphene and the contribution of the SPEs in aluminum to the non-local surface dielectric function, given in Eq.\ (\ref{Es}).
Since a linear expansion of both the MDF-RPA polarization function for graphene, given in Eq.\
(\ref{loss}), and the surface dielectric function for aluminum, obtained from Eqs.\ (\ref{Es}) and (\ref{Lindhard}), is possible for low
frequencies,
we can apply the concept of the friction coefficient $\eta$ defined in Eq.\ (\ref{eta}). We note that the $q$ integral in
Eq.\ (\ref{eta}) takes into account two wavenumber cut-offs: $q<2k_F=2\sqrt{\pi n}$ for graphene, and $q<2K_F\approx$ 3.5
\AA$^{-1}$ for aluminum. Since typical charge carrier densities in graphene have $k_F\ll K_F$, it is interesting to
see how significant graphene's screening ability is in a combined system with a metal. Following
Ref.\cite{Khomyakov_2009}, we take equilibrium density of graphene on aluminum to be $n=n_0=10^{13}$ cm$^{-2}$, and the gap
height to be $h=1$ \AA.

In Fig.\ 5, we show the dependence of the friction coefficient on the distance $z_0$ for a proton moving above free graphene, a
free aluminum surface (which is placed in the plane $z_0=-1$ \AA), and a graphene on aluminum system. We also show the friction
coefficient for a particle moving above free graphene with $n=0$. Although we only consider the case of vanishing damping, we
note that the effects of finite damping can be quite large in both systems \cite{Dedkov_2002,Nunez_1980,Allison_2009}. For the
finite density $n=n_0$, one can see from Fig.\ 5 that while the friction from the free Al surface dominates at short distances,
the friction from free graphene has a longer range and dominates for $z_0>4$ \AA. Furthermore, the friction coefficient of the
combined system is suppressed when compared to both the case of free Al (at short distances) and the case of free graphene (at
long distances). In the combined system with intrinsic graphene, for which there is no direct contribution from graphene's
intra-band SPEs to the friction, one can see that graphene's static screening reduces the friction from low-energy SPEs in
aluminum. Although we do not explore the effects of the density any further, from the $n=0$ and $n=n_0$ cases one can surmise
that the friction of a combined graphene on metal system can be controlled by changing the doping level of graphene externally.

\section{Concluding remarks}

We have provided a simple theoretical model that describes the dielectric coupling of graphene, represented by a 2D gas of
massless Dirac's fermions, and a semi-infinite 3D substrate, represented by a surface response function in a non-local
formulation. The underlying assumption that the electronic band structures of these two systems is not altered by their
proximity has allowed us to study the hybridization of their excitation modes due to Coulomb interaction. An emphasis was placed
on the role of the dynamic response of a substrate in the low-frequency excitations of the combined graphene-substrate system, which
give rise to the energy loss of slow moving charges above graphene. A comparison of the loss function for the graphene on an
insulator system with a HREELS experiment was used as a motivation to explore methods of improving the polarization
function for free graphene in the RPA. After identifying the damping rate and the equilibrium density of charge carriers as the two main
parameters in graphene's polarization function, we calculated the stopping force on a point charge moving parallel to
graphene supported by a SiC substrate, which was described by an empirical dielectric function that included an optical phonon excitation mode
without dispersion. As a result, strong effects from the hybridization of graphene's
$\pi$ plasmon and the substrate's surface phonon were found on the low-velocity stopping force.
As another example, a friction coefficient was calculated for a slow
charge moving above graphene on a metallic substrate by using Lindhard's dielectric function to obtain a surface response function
representing the low-energy single-particle excitations in aluminum in a non-local formulation. In the limit of vanishing
damping rates for both graphene and aluminum, an interesting interplay was found between the low-energy single-particle excitations of both
systems in the friction coefficient. In all cases studied, the low-energy excitation modes in graphene
were found to be strongly coupled with those in the substrate, giving rise to various forms of the stopping force on external
charges that can be of interest to studies such as the chemical reactivity of graphene. Particularly intriguing is the possibility
of affecting the dynamic response of the combined graphene-substrate system by changing the charge density in graphene by
external means, thereby enabling a control of friction at the nano-scale.

\ack

This work was supported by the Natural Sciences and Engineering Research Council of Canada.

\newpage
\begin{figure}
\centering
\includegraphics[width=\textwidth]{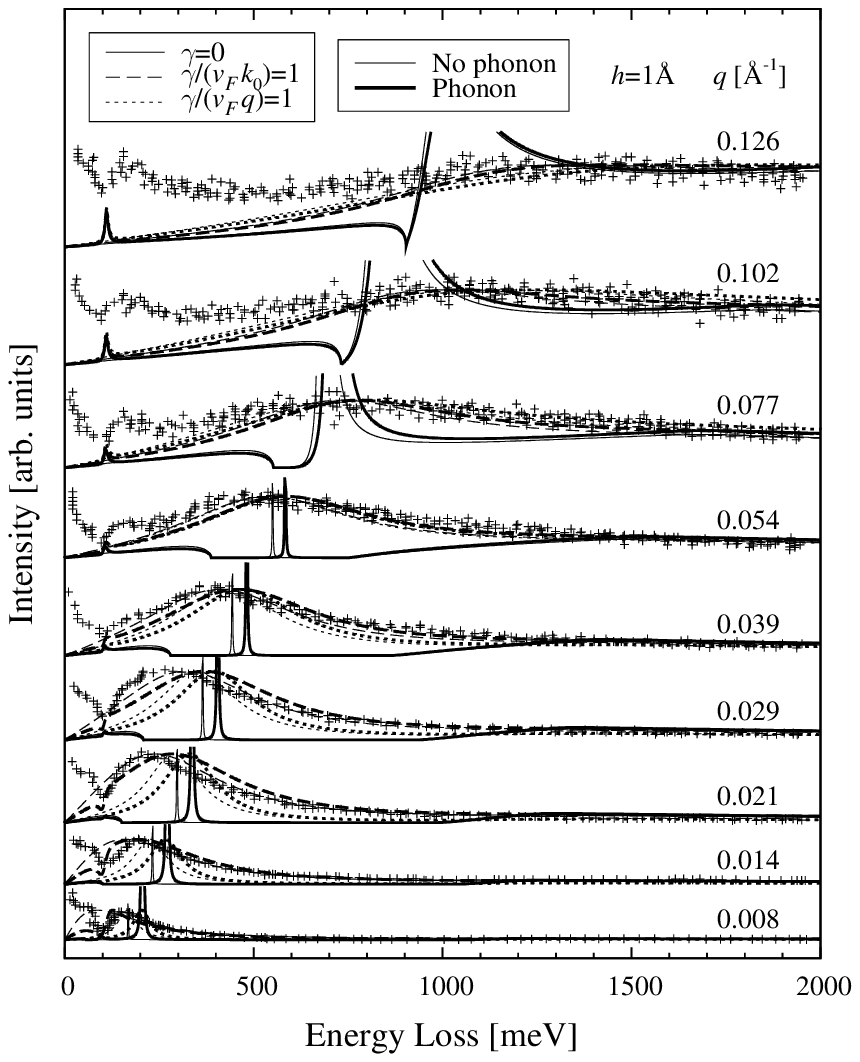}
\caption{The MDF-RPA ($\gamma=0$) and Mermin loss functions ($\gamma \ne 0$) versus the energy loss for graphene with a charge-carrier density
$n=1.9\times 10^{13}$ cm$^{-2}$ supported on a SiC substrate with a gap height $h$ = 1\AA.
Model results for the Mermin loss function include the best-fit constant and linearly dispersing damping rates
$\gamma/(v_Fk_0)=1$ and $\gamma/(v_Fq)=1$, respectively,
where $k_0=\sqrt{\pi n_0}$ and $n_0=10^{13}$ cm$^{-2}$. Thick and thin lines show model results with and without the inclusion of the substrate's TO phonon,
respectively, while symbols show the experimental HREELS data from Ref.\ \cite{Liu_2008}.}
\end{figure}

\begin{figure}
\centering
\includegraphics[width=\textwidth]{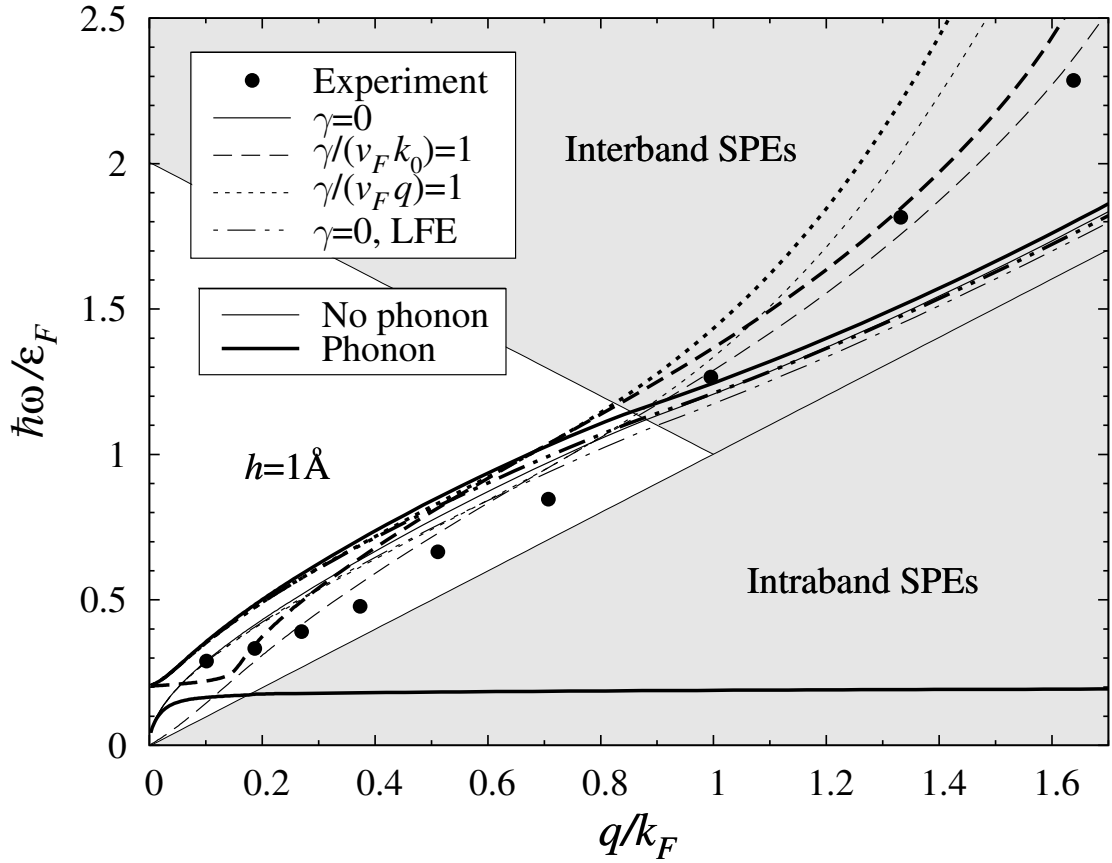}
\caption{The peak positions of the MDF-RPA loss function ($\gamma=0$), the Mermin loss function ($\gamma\ne0$), and the
LFE-corrected MDF-RPA loss function ($\gamma=0$, LFE)
shown as a function of the wavenumber $q/k_F$ for graphene with a
charge-carrier density $n=1.9\times 10^{13}$ cm$^{-2}$ supported on a SiC substrate with a gap
height $h=1$ \AA.
Model results for the Mermin loss function include the best-fit constant and linearly dispersing damping rates
$\gamma/(v_Fk_0)=1$ and $\gamma/(v_Fq)=1$, respectively,
where $k_0=\sqrt{\pi n_0}$ and $n_0=10^{13}$ cm$^{-2}$. Thick and thin lines show model results with and without the inclusion of the substrate's TO phonon,
respectively, while filled circles show the experimental HREELS data from Ref.\ \cite{Liu_2008}.}
\end{figure}

\begin{figure}
\centering
\includegraphics[width=\textwidth]{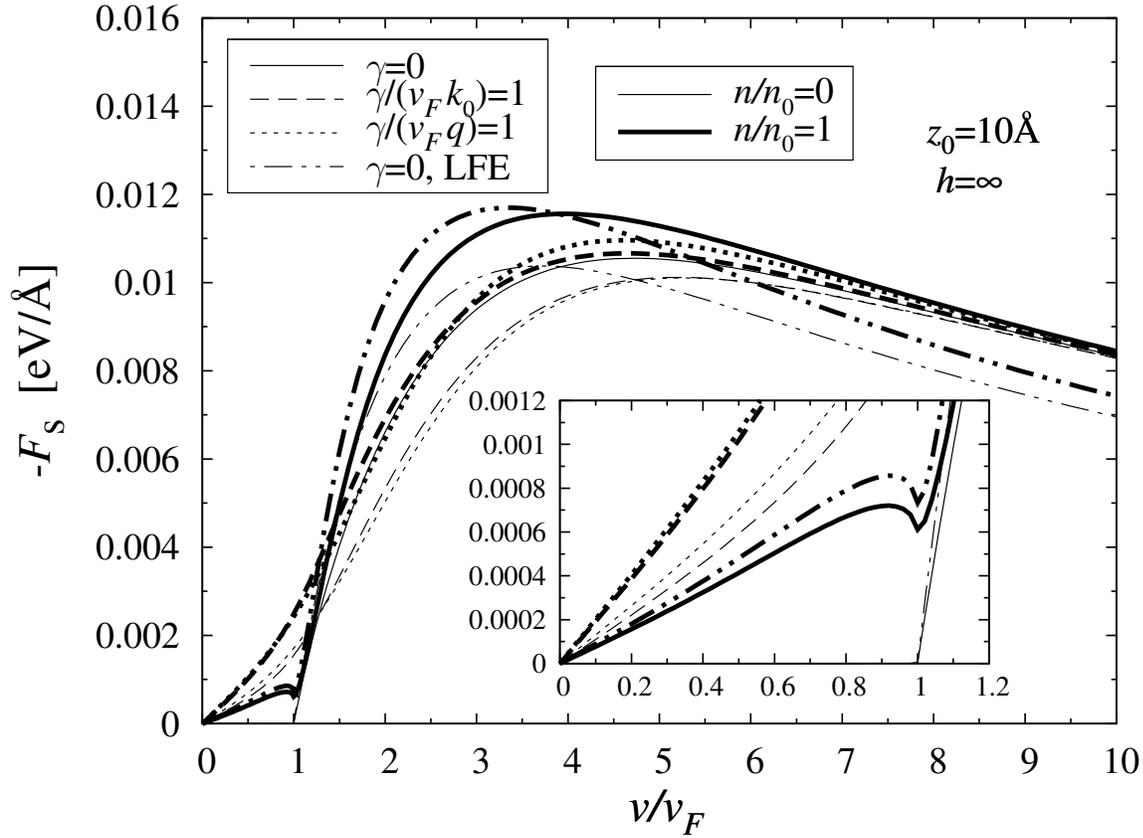}
\caption{ The stopping force from the MDF-RPA loss function ($\gamma=0$), the Mermin loss function ($\gamma\ne0$), and the
LFE-corrected MDF-RPA loss function ($\gamma=0$, LFE)
shown as a function of the reduced speed $v/v_F$ of a proton ($Z=1$) moving at a distance $z_0$ = 10 \AA\ above free graphene ($h=\infty$).
Results for the Mermin loss function include the best-fit constant and linearly dispersing damping rates
$\gamma/(v_Fk_0)=1$ and $\gamma/(v_Fq)=1$, respectively,
where $k_0=\sqrt{\pi n_0}$ and $n_0=10^{13}$ cm$^{-2}$.
Thin and thick lines represent the case of intrinsic graphene
and graphene with a reduced charge carrier density $n/n_0=1$, respectively. }
\end{figure}

\begin{figure}
\centering
\includegraphics[width=\textwidth]{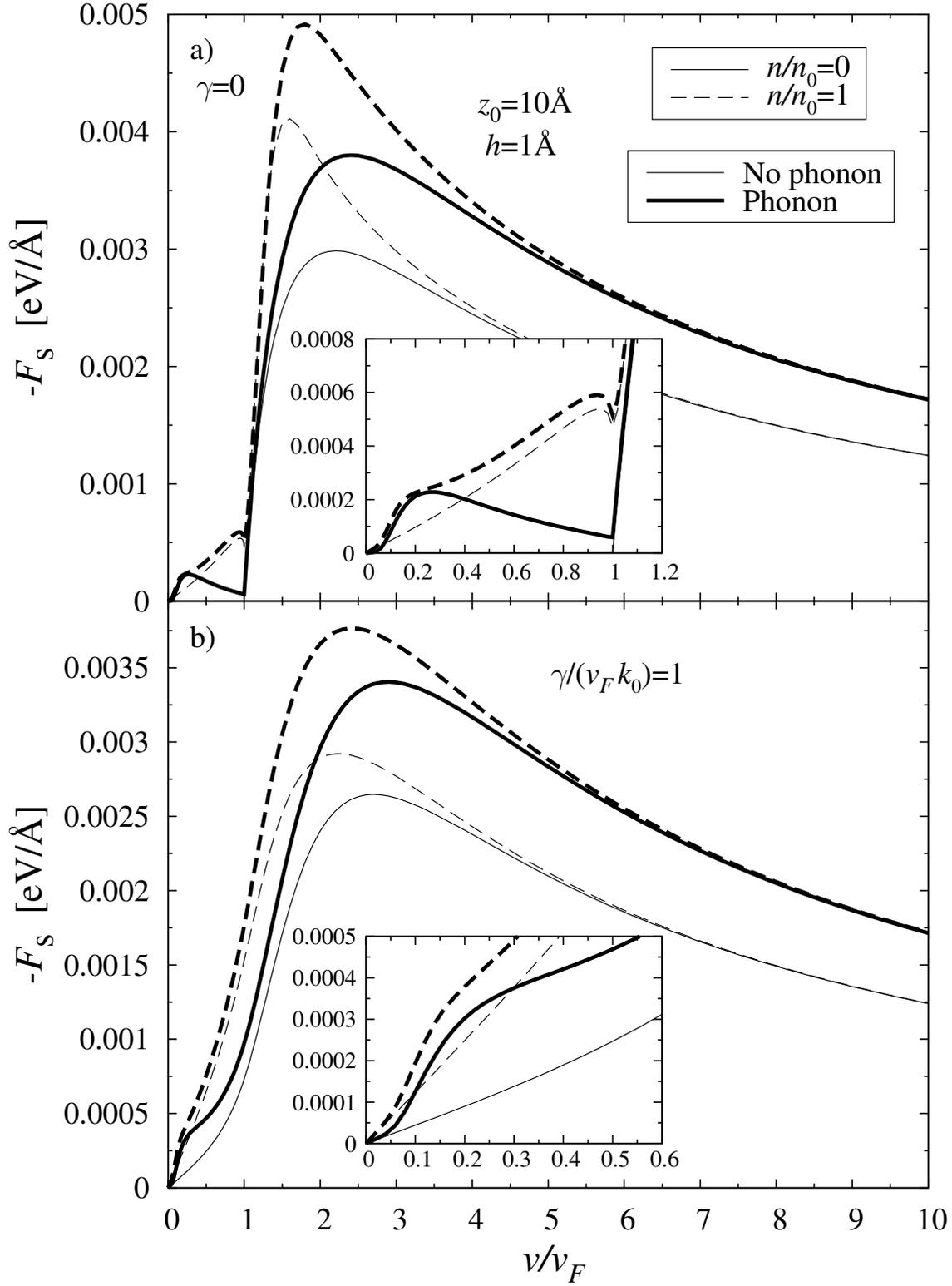}
\caption{ The stopping force from (a) the MDF-RPA loss function ($\gamma=0$) and (b) the Mermin loss function with $\gamma/(v_Fk_0)=1$,
where $k_0=\sqrt{\pi n_0}$ and $n_0=10^{13}$ cm$^{-2}$,
shown as a function of the reduced speed $v/v_F$ of a proton ($Z=1$) moving at a distance $z_0$ = 10 \AA\ above graphene
supported on a SiC substrate with a gap height $h=1$ \AA.
Results are shown for intrinsic graphene (solid lines) and graphene with a reduced charge carrier density $n/n_0=1$ (dashed lines), as well as for
the model with (thick lines) and without (thin lines) the inclusion of the substrate's TO phonon.}
\end{figure}

\begin{figure}
\centering
\includegraphics[width=\textwidth]{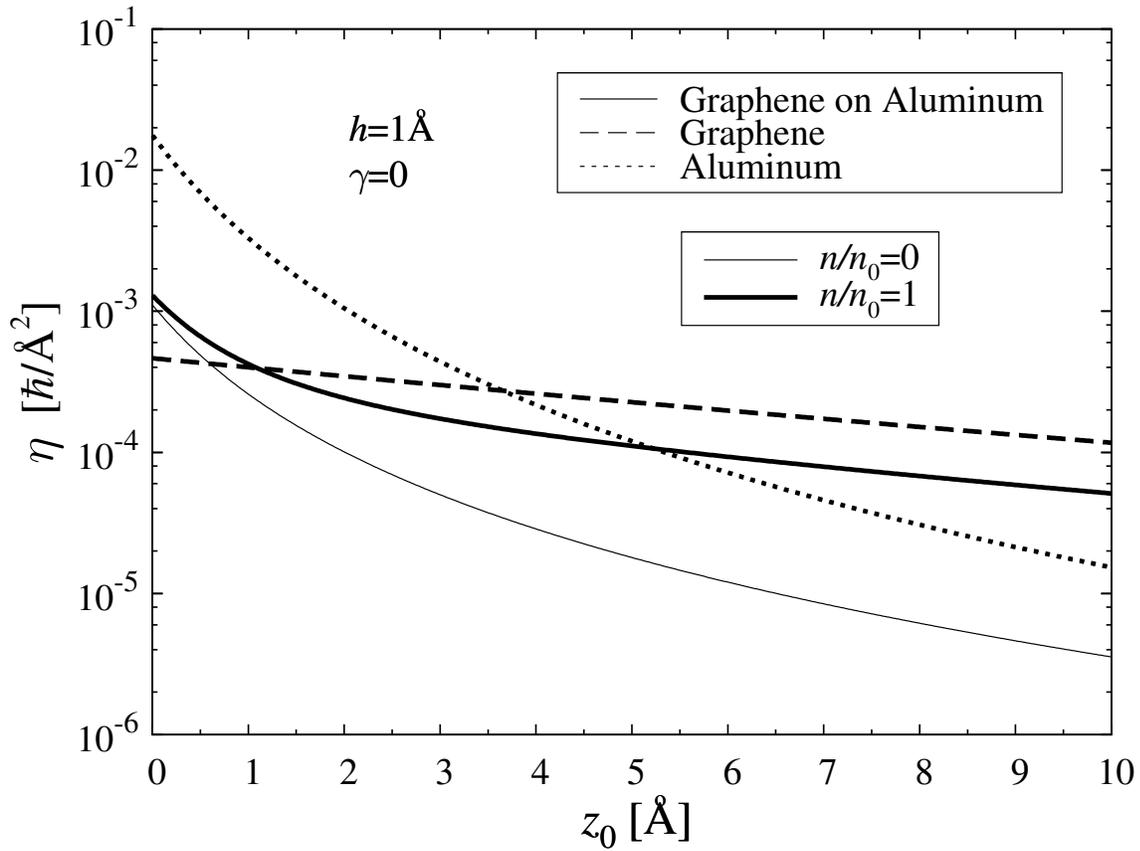}
\caption{The friction coefficient $\eta$ shown as a
function of the distance $z_0$ of a proton ($Z=1$) moving above graphene on an aluminum substrate (solid lines), free graphene (dashed lines), and
a free aluminum substrate (dotted lines).
Where appropriate, thin and thick lines represent the case of intrinsic graphene
and graphene with a reduced charge carrier density $n/n_0=1$, where $n_0=10^{13}$ cm$^{-2}$, respectively. }
\end{figure}

\end{document}